*Article*

# Entropy-based machine learning model for diagnosis and monitoring of Parkinson's Disease in smart IoT environment

Maksim Belyaev [1,*], Murugappan Murugappan [2,3,4], Andrei Velichko [1] and Dmitry Korzun [5]

[1] Institute of Physics and Technology, Petrozavodsk State University, 185910 Petrozavodsk, Russia
[2] Intelligent Signal Processing (ISP) Research Lab, Department of Electronics and Communication Engineering, Kuwait College of Science and Technology, Block 4, Doha, Kuwait.
[3] Department of Electronics and Communication Engineering, Faculty of Engineering, Vels Institute of Sciences, Technology, and Advanced Studies, Chennai, India
[4] Centre of Excellence for Unmanned Aerial Systems (CoEUAS), Universiti Malaysia Perlis, 02600, Arau, Perlis, Malaysia.
[5] Department of Computer Science, Institute of Mathematics and Information Technology, Petrozavodsk State University, 185910 Petrozavodsk, Russia
*Correspondence: biomax89@yandex.ru;

**Abstract:** The study presents the concept of a computationally efficient machine learning (ML) model for diagnosing and monitoring Parkinson's disease (PD) in an Internet of Things (IoT) environment using rest-state EEG signals (rs-EEG). We computed different types of entropy from EEG signals and found that Fuzzy Entropy performed the best in diagnosing and monitoring PD using rs-EEG. We also investigated different combinations of signal frequency ranges and EEG channels to accurately diagnose PD. Finally, with a fewer number of features (11 features), we achieved a maximum classification accuracy ($A_{RKF}$) of ~99.9%. The most prominent frequency range of EEG signals has been identified, and we have found that high classification accuracy depends on low-frequency signal components (0-4 Hz). Moreover, the most informative signals were mainly received from the right hemisphere of the head (F8, P8, T8, FC6). Furthermore, we assessed the accuracy of the diagnosis of PD using three different lengths of EEG data (150-1000 samples). Because the computational complexity is reduced by reducing the input data. As a result, we have achieved a maximum mean accuracy of 99.9% for a sample length ($L_{EEG}$) of 1000 (~7.8 seconds), 98.2% with a $L_{EEG}$ of 800 (~6.2 seconds), and 79.3% for $L_{EEG}$ = 150 (~1.2 seconds). By reducing the number of features and segment lengths, the computational cost of classification can be reduced. Lower-performance smart ML sensors can be used in IoT environments for enhances human resilience to PD.

## 1. Introduction

By 2030, experts predict that every sixth person on earth will be over 60 years of age due to an increasing life expectancy [1]. It is estimated that 1.4 billion people will be over 60 by 2050, and this number is expected to double by then. Age-related neurodegenerative diseases are a major risk factor for mortality and morbidity caused by neurodegenerative diseases [2–4]. The symptoms of neurodegenerative disease may begin as early as middle age [5], followed by overt signs and symptoms. By diagnosing and treating patients early, irreversible damage to the nervous system can be reduced, improving their quality of life and length of life.

In addition to diagnostics, a personalized approach to neurodegenerative disease treatment using IoT-enabled environments is essential to improving patients' quality of life [5,6], such as smart homes [7], smart retail spaces [8], and smart healthcare [7]. The Healthcare Internet of Things (H-IoT) [10] is also known as IoMT [9] and is one of the most efficient tools for this purpose. An IoMT network is capable of continuously monitoring physiological parameter changes in humans by using machine learning (ML) models trained on smart sensors [11–13]. Physiological or biomedical sensors that are placed on the patient's body (wearable sensors) measure different types of physiological responses, including heart rate, blood pressure, skin electrical conductivity, oxygen saturation, heart electrical activity, electroencephalograms (EEGs), etc. [14]. Additionally, some sensors can be placed in the room where the patient is located to monitor their movement patterns, gait, physical activity, etc. [15,16]. Smart IoT environments are capable of collecting and analyzing a wide array of information in real-time using smart ML sensors, then presenting the results to both the patient and the attending physician through remote, authorized access to system data (Figure 1).

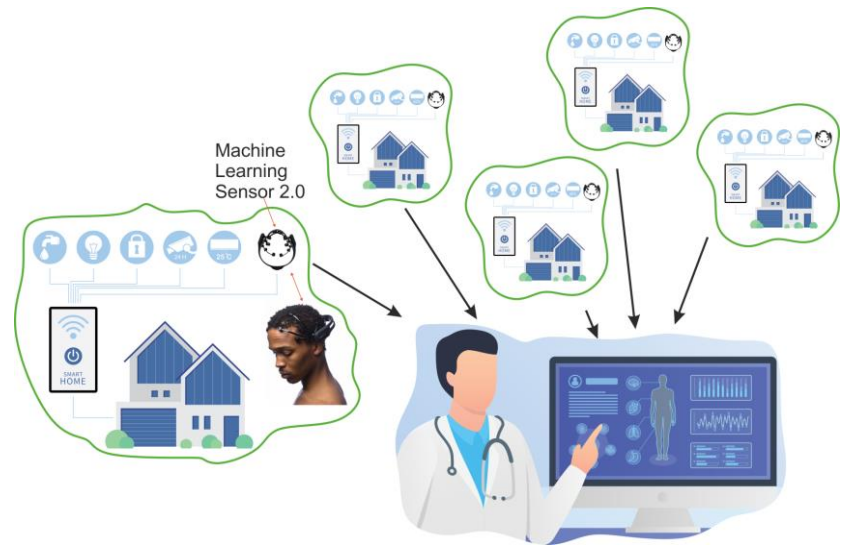

**Figure 1.** The concept of a smart IoT environment that can continuously monitor Parkinson's disease patients.

The latter is especially important if the treatment takes place at home rather than in a medical facility [17]. An attending physician can intervene quickly if a patient's condition deteriorates, which the patient himself/herself may not be aware of due to the deterioration of cognitive functions. This approach enhances human resilience to PD. In addition to transforming the hardware designs of traditional sensor systems using ML techniques, artificial intelligence sensors (or smart sensors) can also be designed holistically based on ML methods [18] and machine learning algorithms [19, 20]. A further development of the ML sensor paradigm was made by Warden et al. [12] and Matthew Stewart [11], where the authors introduced the terms Sensors 1.0 and Sensors 2.0. Sensors 2.0 involve both a sensor and a machine learning module integrated into one device. This study presents an entropy-based machine learning model using EEG sensors. This ML model diagnoses and monitors Parkinson's disease. The sensor would be attributed to Sensors 2.0 devices if implemented on the real device (wireless headset) (Figure 1). EEG signals will be input into the model, and the output will be the degree of disease development. This may be part of a smart IoT environment for patient health monitoring.

Presently, there are a number of methods available for detecting neurodegenerative diseases. Using the data collected from question assessments [21], blood biomarkers [22], eye tracking parameters [23], kinematic gait parameters [24], electroencephalogram (EEG) [25], and other tests, conclusions are drawn regarding the presence or high risk of developing PD.

Among the disease indicators presented, EEG is one of the most promising because of its non-invasive nature, wide distribution, low cost, and ability to be integrated into the Internet of Things (IoT) [26]. In addition to diagnosing diseases, the use of portable personal devices for recording EEG can be used to continuously monitor the patient's current condition and the effectiveness of the selected treatment method outside of the medical institution through the development of intelligent devices with IoT technology. Usually, EEG signals can be analyzed in time, frequency, and time-frequency domains [28]. A time-frequency analysis can be done by applying a short-time Fourier transform [29] and wavelet transform [30] to examine the local temporal effects that occur under specific bands of EEG frequencies. Generally, five different frequency bands are investigated in EEG signals depending on their application. These bands are delta (0-4 Hz), theta (4-8 Hz), alpha (8-13 Hz), beta (13-30 Hz), and gamma (30-49 Hz) [31].

The electrical signals from the brain are highly non-stationary and complex. They are susceptible to disturbances caused by external and internal noises. To characterize EEG signal behavior in each class, different types of features are needed, such as statistical, spectral, and entropy characteristics [32–34]. Various machine learning methods are then used to determine the presence and type of neurodegenerative diseases [35–40]: artificial neural networks (ANN), probabilistic neural networks (PNN), support vector machines (SVM), neural networks (NN), including deep learning (DNN), decision trees (DT), random forests (RF), Bayesian models (NB), k-nearest neighbor method (KNN), etc.

Rest-state EEG signals can be used to detect a variety of neurodegenerative diseases, including Alzheimer's disease [41], Parkinson's disease (PD) [37-40,42-44], frontotemporal dementia [45], dementia with Lewy bodies [46], and epilepsy [47]. In the field of neurodegenerative diseases, Parkinson's disease is one of the most studied. As a result of Parkinson's disease, a person will have impaired motor functions (slowness of movement, tremors, rigidity, and loss of balance) and impaired non-motor functions (decreased cognitive functions, mental disorders, sleep disturbances, pain, and sensory problems) [48]. As part of PD diagnostics, the EEG signal of the patient is compared to the EEG signal of a healthy control group. Depending on the metric, changes can be determined either by comparing signals directly [43] or by quantifying them using entropy metrics, spectral power metrics, cross-correlation metrics, statistical values, etc. [37–40,42,44]. In [43], convolutional neural networks are used to classify PD using a deep learning approach, in which the elements of the filtered signal are fed into the neural network as input. A classification accuracy of 88.2% was achieved by this approach. However, using calculated features usually yields better results. A principal component analysis of the filtered signal, correlation coefficients, and linear predictive coefficients is used to calculate features for the SVM classifier [44] that achieved a maximum mean classification rate of 99.1% in diagnosing PD. Due to the fact that EEG changes can occur at certain frequency ranges (corresponding to alpha, beta, gamma, theta, and delta waves), translating the signal from the time domain (Fourier transform) or frequency-time domain (wavelet transform) is a common method of analyzing an EEG signal. According to [39], spectral features (such as wavelet coherence and relative wavelet energy) were used to detect PD-related dementia, AD, and control groups. Spectral energy differences were found between the control group and the rest of the patients both at low and high frequencies. They were able to determine PD with an accuracy of 79.1% and AD with an accuracy of 81.2% using linear discriminant analysis. By using the tunable Q wavelet transform, statistical signal metrics were extracted from the frequency subbands of the rest-state EEG signal (minimum, Hurst exponent, Higuchi fractal dimension, Hjorth complexity, mobility). There are four types of classifiers used to distinguish PD from healthy controls (ANN, SVM, KNN, and RF) based on the above-mentioned statistical features. As a result, the mean classification accuracy for healthy controls and PD patients (with or without medical treatment) was 96.1% and 97.7%, respectively.

Several works have focused on the difference in entropy of signals in different frequency ranges between patients with PD and the control group [37,40,42]. According to [42], relative spectral powers and wavelet packet entropy were used to identify PD. Alt-hough entropy features allow better separation of two classes, relative spectral power (especially in the beta band) can also be useful. Higher-order spectral features, like bispectral entropies and mean magnitude, are used for PD diagnosis [37] based on five different types of classification algorithms such as DT, KNN, NB, PNN, and SVM. The SVM classifier reported a maximum mean accuracy of 99.6% compared to other classifiers in diagnosing PD. The authors in [40] used KNN and SVM classifiers to diagnose PD based on energy and entropy features extracted from reconstructed wavelet signals. Accordingly, KNN and SVM classifiers achieve 99.5% and 99.9% mean accuracy, respectively.

Although the presented results prove a high classification accuracy (more than 99%), most of the approaches used to calculate features are limited. Also, the hyperparameters used when calculating the entropy can significantly affect the calculation result. To obtain high accuracy, a number of studies have used many features [39,40,42,43], which complicates the implementation of the methods in low-performance IoT devices. This paper attempts to address these deficiencies by comparing various entropy methods, carefully selecting their parameters, and analyzing EEG signal frequency ranges for diagnosing PD. By analyzing EEG data collected from Normal Control (NC) and Parkinson's Disease (PD) patients using Wireless Emotiv's Epoc headsets, we have developed a novel method for detecting Parkinson's Disease.

The major contributions of the paper are:

- In this study, we explored the most effective entropy method for calculating EEG entropy features. This method enables us to separate the EEG signals between the PD and NC with the highest accuracy. In the diagnosis and monitoring of Parkinson's disease, fuzzy entropy has shown the best properties among other types of entropy.

- By identifying the most prominent EEG frequency ranges and channels (electrodes), as well as their combinations, we were able to diagnose the PD with less computational complexity than other state-of-the-art studies.

- We propose a method to monitor a patient's condition based on an entropy value. The method allows the creation of an ML sensor model for Parkinson's disease detection and monitoring. With only 11 features, 99.9% classification accuracy is achieved.

- In this paper, we propose the concept of an IoT environment that monitors PD patients' health status using a machine learning model. By reducing the number of features, classification costs can be reduced, and low-performance sensors can be used to deploy an IoT system for health monitoring.

Following is an outline of the remainder of the paper. Section 2 provides an overview of the datasets used, proposed methods, and performance metrics. A comparison of classification accuracy using different EEG channels and frequency bands is presented in Section 3. In section 4, we present the concept of an IoT system that monitors the health status of the patient. The study's conclusion and future directions are outlined in Section 5.

**2. Materials and Methods**

*2.1 Dataset*

This study was conducted using an EEG dataset consisting of 20 patients with Parkinson's disease and 20 age-matched normal control subjects without a history of psycho-logical disorders or neurological disorders. This dataset was collected at the Hospital Universiti Kebangsaan Malaysia in Malaysia. The entire data acquisition protocol at the Hospital Universiti Kebangsaan Malaysia has been approved by the Institutional Ethical Review Board Committee as part of the hospital's ethical review process. An Emotiv EPOC wireless headset with a total of 14 channels (Figure 2a) was used for recording EEG signals from both NCs and PDs in the rest-state condition with the eyes closed for a period of 5 minutes during this study. In accordance with the international standard 10-20 system, the 14 channels (AF3, F7, F3, FC5, T7, P7, O1, O2, P8, T8, FC6, F4, F8, AF4) are placed on the subject's scalp (Figure 2b). With a sampling rate of 128 Hz, the data collected for each of the channels was converted into digital signals. Using the Hoehn and Yahr scales, a total of seven patients were classified as having Parkinson's disease stage III, eleven patients as having Parkinson's disease stage II, and two patients as having Parkinson's disease stage I. A complete description of the dataset, acquisition, and pre-processing of the dataset can be found in [49–51]. Since the number of patients is relatively small, each EEG record was divided into 5 non-overlapping segments, each of which represented an independent observation within the framework of this study. The duration of all segments was the same and amounted to ~7.8 seconds (1000 counts).

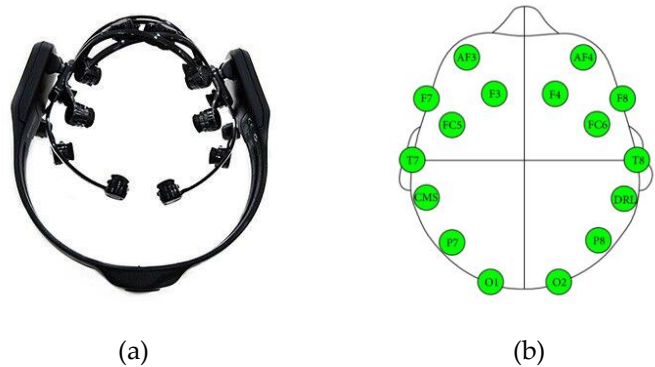

(a) (b)

**Figure 2.** Emotiv EPOC wireless headset (a). Location of the electrodes on the head (b) [52]

*2.2 Signal preprocessing*

Considering the wide spectral range of EEG signals (0-64 Hz), and the fact that most brain activity information is contained in relatively narrow frequency subranges [53,54], it is possible that the entropy of the original signal gives a poor indication of separation capability. Filtering the initial data and decomposing the signal into separate frequencies using the wavelet transform can increase EEG signals' information content. A fifth-order Butterworth filter with a cut-off frequency of 0.5-32 Hz is applied to all acquired signals to remove low- and high-frequency noise, while amplitude thresholding of ±85μV is applied to remove artifacts (eye blinking, eyeball rotation, and eye movements) during the acquisition process. A discrete wavelet transform (DWT) was performed on the signal using the db4 wavelet [55]. After decomposing into wavelet approximation coefficients (cA1-cA4) and detail (cD1-cD4), each of them was utilized to reconstruct the signals, with each signal (cA1-cA4 and cD1-cD4) being reconstructed with only one of the coefficients. A similar method was used in [40], however, a different frequency band was chosen.

Using the original dataset, 9 variants of different signal types were obtained:

- Original (O) signal; Frequency ranges: (0-64Hz);
- Signals reconstructed based on approximation coefficients (cA1-cA4); Frequency ranges: (cA1 (0-32Hz), cA2 (0-16 Hz), cA3 (0-8 Hz), cA4 (0-4 Hz));

- Signals reconstructed based on detail coefficients (cD1-cD4); Frequency ranges: (cD1 (32-64Hz), cD2 (16-32Hz), cD3 (8-16 Hz), cD4 (4-8 Hz));

*2.3 Feature generation*

The entropy features were calculated from the EEG signals after applying DWT and concatenated to form the feature vector for each class (NC and PD). Later, these feature vectors are used in classifying patients using different machine-learning methods. This entropy model comprises several features such as Singular Value Decomposition Entropy (SVDEn), Permutation Entropy (PermEn), Sample Entropy (SampEn), Cosine Similarity Entropy (CoSiEn), Fuzzy Entropy (FuzzyEn), Phase Entropy (PhaseEn), and Attention Entropy (AttnEn). A method for calculating entropy was implemented using the EntropyHub [56] software package, except for SVDEn and PermEn. The Antropy [57] software package was used to calculate SVDEn and PermEn. The range of hyperparameters used for computing each type of entropy is shown in Table 1. There are no hyperparameters associated with the AttnEn.

**Table 1.** Ranges of parameters used to create the entropy features

| Entropy name | Parameter range |
| --- | --- |
| SVDEn | order m = 2..10 |
| PermEn | order m = 2..10 |
| SampEn | order m = 1..3, tolerance r = 0.05..0.5×std |
| CoSiEn | order m = 2..3, tolerance r = 0.05..0.5 |
| FuzzyEn | order m = 1..2, tolerance r = 0.05..0.5×std, exponent membership function of order $r_2$ = 1..5 |
| PhaseEn | K = 2..10 |
| AttnEn | No parameters |

*2.4 Assessment of classification accuracy*

The accuracy of classifications was assessed using support vector classifiers (SVCs) implemented using scikit-learn. Two stages were involved in the classification accuracy assessment. In the first step, hyperparameters were selected by means of repeated K-fold cross-validation (RKF) [58]. This was done by dividing the estimated datasets into K= 10 blocks in various ways with *N*= 10. For each of the *N* variants of partitions, the K-blocks were filled with different samples, resulting in a uniform distribution of classes. Sets of samples were created based on K-blocks for training and validating the classifier, with each K-block being validated once and the remaining K-1= 9 times being used in training.

The classifier hyperparameters were then selected at which maximum average accuracy was achieved on the validation set. K-block cross-validation allows the selection of hyperparameter values that do not require retraining of the model because many training and validation sets are used. Due to the optimization of hyperparameters on a fixed set of samples, it is possible that the average cross-validation accuracy is too optimistic. Consequently, after determining the optimal hyperparameters, the next step was taken. During the second stage, optimal values of hyperparameters were used and cross-validation was performed on other *N*= 30 partitions divided into K= 10 blocks, which was different from the first stage. The accuracy of classification was measured based on the average $A_{RKF}$ accuracy across the new partitions.

**3. Experimental Results and Discussion**

In this section, we present the results of assessing classification accuracy using all features, one signal type, all channels, one channel, and one feature.

*3.1 Classification accuracy using one method for calculating the entropy*

In both NC and PD, the entropy feature was computed using all nine types of input EEG data (original signal and eight reconstructed signals based on detail and approximation coefficients) across 14 channels (126 features in total). A model is developed to categorize NC and PD based on the features extracted from NC and PD pairs. Based on PermEn, SampEn, CoSiEn, FuzzyEn, PhaseEn, BubbleEn, and SVDEn, Figure 3 shows the accuracy of classification ($A_{RKF}$) of each entropy feature with different hyperparameters. These entropy features have been computed with varying hyperparameter values in this study. Using five non-overlapping segments of 40 subjects (20 PDs and 20 NCs), we extracted the entropy features from 200 datasets. In this task, the optimal parameters for each of the entropy calculations were determined.

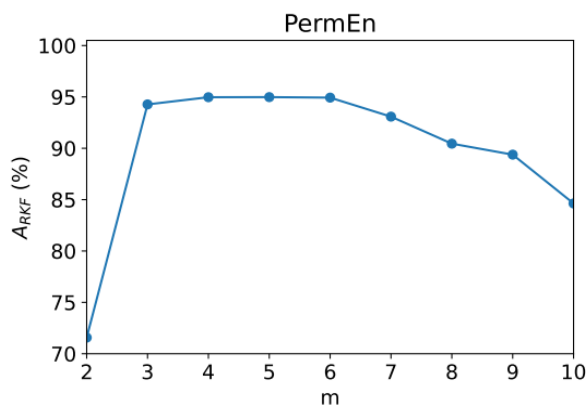

(a)

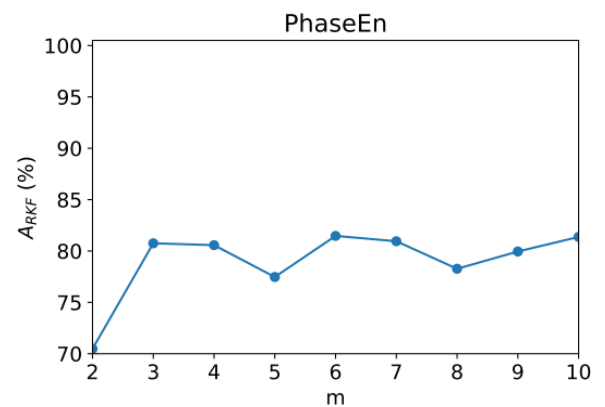

(b)

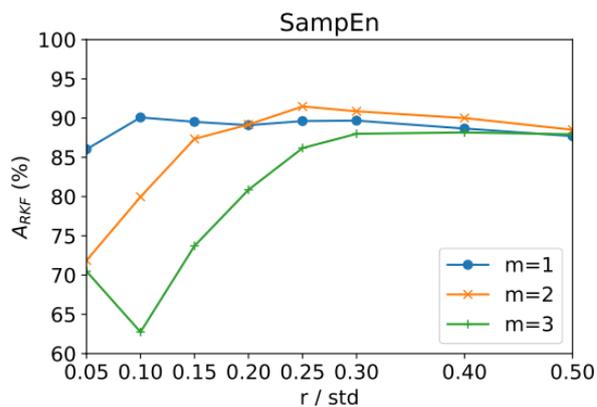

(c)

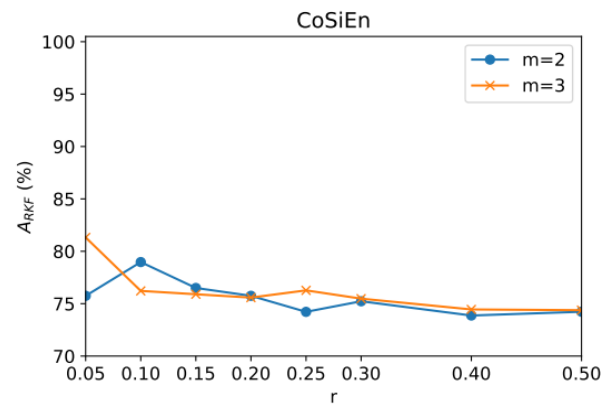

(d)

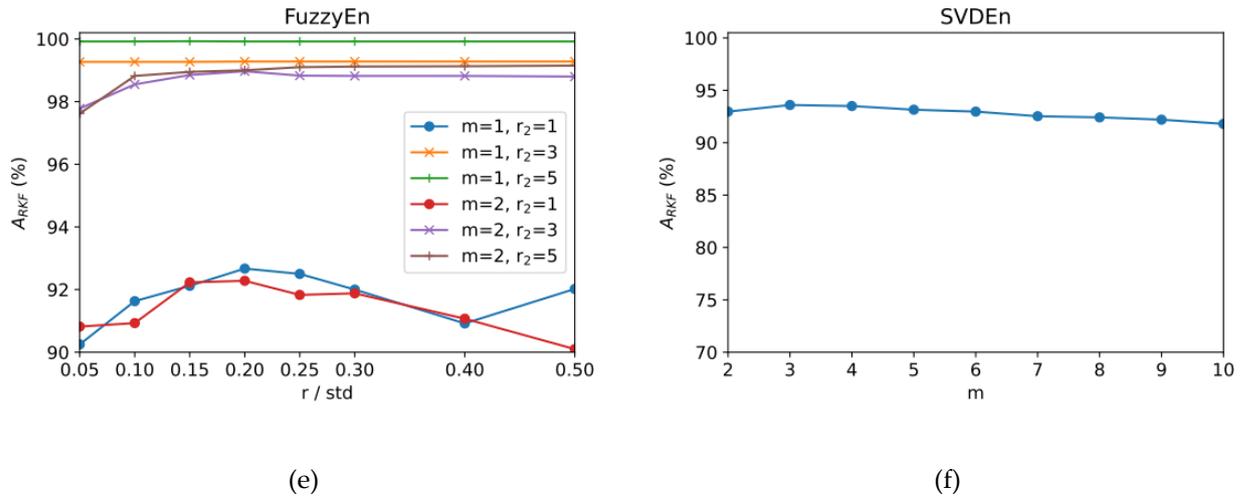

**Figure 3.** Dependence of $A_{RKF}$ on entropy parameters using all 126 features for: PermEn (a), PhaseEn (b), SampEn (c), CoSiEn (d), FuzzyEn (e), SVDEn (f).

The best classification result $A_{RKF}$ = 99.9% was demonstrated for FuzzyEn with parameters ($m$ = 1, $r$ = 0.15×std, $r_2$ = 5). The influence of the parameter $r$ in this case is insignificant. It has been observed that the $A_{RKF}$ value increases as the parameter $r_2$ increases from 1 to 5. The next most accurate entropy method was AttnEn ($A_{RKF}$ = 97.9%). This method has no hyperparameters. Acceptable accuracy was achieved for PermEn ($A_{RKF}$ = 95% for $m$ = 5) and SVDEn ($A_{RKF}$ = 93.6% for $m$ = 3). Both curves have a maximum at intermediate values of the parameter m. The worst results were obtained using the SampEn ($A_{RKF}$ = 91.5% for $m$ = 2, $r$ = 0.25×std), PhaseEn ($A_{RKF}$ = 81.5% for $K$ = 6), and CoSiEn ($A_{RKF}$ = 81.3% for $m$ = 3, $r$ = 0.05) methods.

*3.2 Classification accuracy using one type of signal*

Furthermore, we wish to identify which type of EEG data is most effective among the nine types of data, as described in Section 2.2, based on different entropy measures. Through this investigation, the computational complexity (memory and computation time) of the proposed PD diagnosis system can be reduced. This section presents the results of calculating the average accuracy ($A_{RKF}$) using each type of nine signals (O, cA1-cA4, cD1-cD4) for each of the 14 channels (14 features in total). The values of the optimal entropy parameters correspond to those presented in Section 3.1. Figure 4 shows the dependence of $A_{RKF}$ on the type of signal.

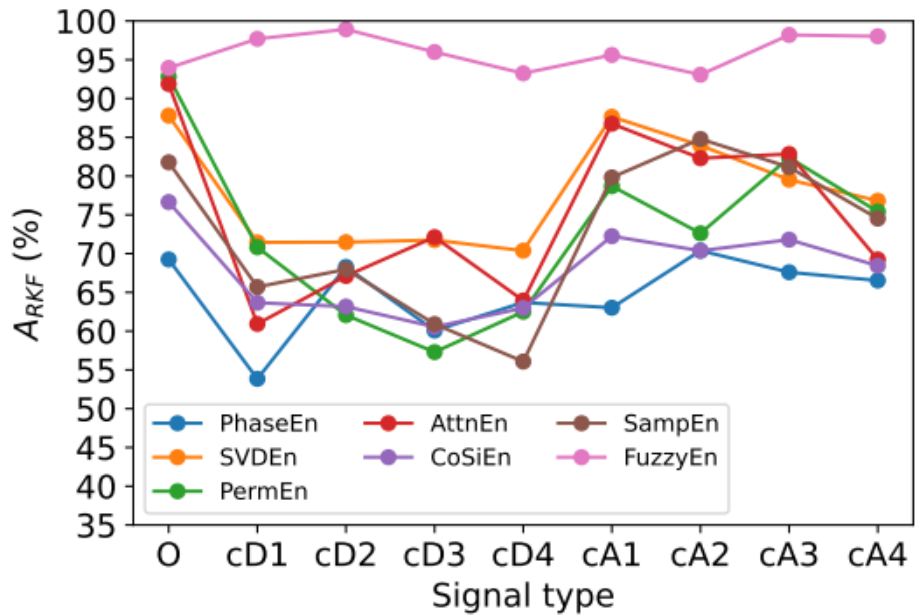

**Figure 4.** Dependence of $A_{RKF}$ on signal type for different entropy calculation methods: PhaseEn ($K$ = 6), SVDEn ($m$ = 3), PermEn ($m$ = 5), AttnEn, CoSiEn ($m$ = 3, $r$ = 0.05), SampEn ($m$ = 2, $r$ = 0.25×std), FuzzyEn ($m$ = 1, $r$ = 0.15×std, $r_2$ = 5).

According to the experimental results, FuzzyEn has higher accuracy than other types of entropy features for all types of signals. According to the presented data, it can be noted that the use of only one type of signal (14 features) generally reduces the accuracy of the $A_{RKF}$ classification compared to using all 126 features. When using FuzzyEn, the $A_{RKF}$ value had high values for the signals: cD2 ($A_{RKF}$ = 98.9%), cA3 ($A_{RKF}$ = 98.2%), cA4($A_{RKF}$ = 98%). For other entropies, high $A_{RKF}$ values were observed for signals O, cA1, cA2, cA3, cA4. Perhaps this is due to the presence in these signals of a low-frequency component in the range from 0 to 4 Hz: O (0-64Hz), cA1 (0-32Hz), cA2 (0-16 Hz), cA3 (0-8 Hz), cA4 (0-4 Hz), while cD1 (32-64Hz), cD2 (16-32Hz), cD3 (8-16 Hz), cD4 (4-8 Hz) signals contain higher frequency components. Low-frequency rhythms (delta and theta) are usually prominent while the eye is closed and in a resting state compared to waking and alert states (while the eye is open and focused). People with neurological disorders, particularly those with delta and theta rhythms, tend to have these rhythms dominate more than healthy individuals. Due to this, low-frequency rhythms (alpha to gamma) are more accurate in diagnosing Parkinson's disease than high-frequency rhythms.

The decrease in accuracy when using only one type of signal is quite significant: the classification error $E_{RKF}$ = 1- $A_{RKF}$ increased by 11 times compared to the result when using all features (Section 3.1). Thus, the use of one frequency range is not enough to achieve the maximum classification accuracy $A_{RKF}$ = 99.9%.

*3.3 Classification accuracy using a single channel*

In this section, we present the results of the $A_{RKF}$ classification accuracy using all 9 signal types (9 features in total) corresponding to one of the 14 channels (AF3, F7, F3, FC5, T7, P7, O1, O2, P8, T8, FC6, F4, F8, AF4). The values of the optimal entropy parameters are specified in 3.1. In Figure 5, the $A_{RKF}$ is shown in relation to the channel number.

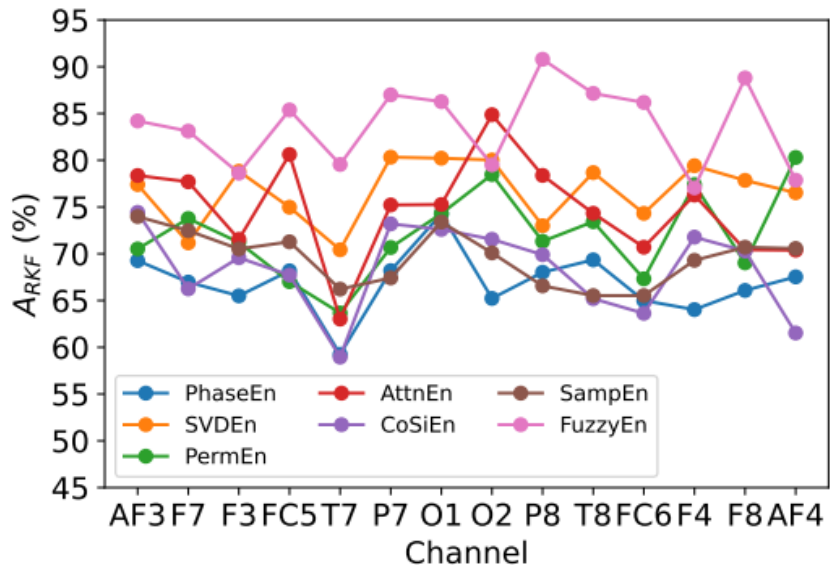

**Figure 5.** Dependence of $A_{RKF}$ on the channel number for different entropy calculation methods: PhaseEn (K = 6), SVDEn (*m* = 3), PermEn (*m* = 5), AttnEn, CoSiEn (*m* = 3, *r* = 0.05), SampEn (*m* = 2, *r* = 0.25×std), FuzzyEn (*m* = 1, *r* = 0.15×std, $r_2$ = 5).

Analyzing the results presented in Figure 5, it can be noted that the highest $A_{RKF}$ value for most channels was obtained using FuzzyEn for the P8 ($A_{RKF}$ = 90.8%) and F8 ($A_{RKF}$ = 88.8%) channels. It is not possible to find pronounced dependencies that are repeated for all entropies. The classification accuracy when using one channel is significantly reduced compared to the results when using all channels: the minimum classification error $E_{RKF}$ increases by ~ 8 times when using one channel and one type of signal (section 3.2) and 92 times when using all signals and all channels (section 3.1). This suggests the need to use multi-channel EEG measurement devices to maximize accuracy.

*3.4 Classification accuracy when using one feature*

In sections 3.2 and 3.3, reduced datasets with 14 (one signal type) and 9 (one channel) features were used, however Figures 4 and 5 show that the classification accuracy varies significantly across different channels and signal types (frequency bands). At the same time, when analyzing by these two criteria, we cannot determine the most informative combinations of channels and frequency ranges.

This section presents the results of using one feature (one type of signal for one channel). In this case, the FuzzyEn method will be used, which gave the best accuracy estimate in Sections 3.2 and 3.3, with the parameters *m* = 1, *r* = 0.15×std, $r_2$ = 5. The graphs are grouped by signal types and were divided into 2 groups:

1. Group 1 consists of signals based on the detail wavelet coefficients: cD1 (32-64Hz), and cD2 (16-32Hz), cD3 (8-16 Hz) and cD4 (4-8 Hz);

2. Group 2 consists of the original signal and signals based on the approximation wavelet coefficients: O (0-64Hz), cA1 (0-32Hz), cA2 (0-16 Hz), cA3 (0-8 Hz) and cA4 (0-4 Hz).

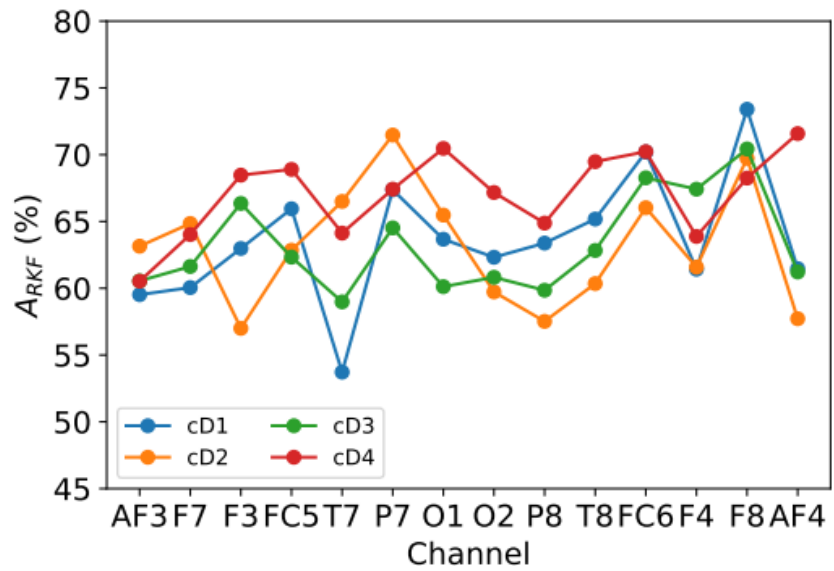

(a)

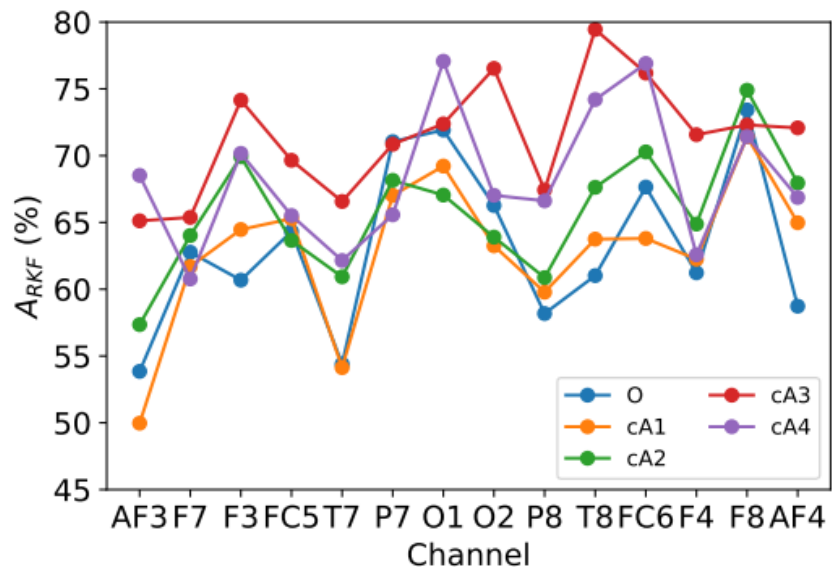

(b)

**Figure 6.** Dependence of $A_{RKF}$ on the channel number for the FuzzyEn method ($m = 1$, $r = 0.15 \times$std, $r_2 = 5$), grouped by signal types: (a) cD1, cD2, cD3, cD4; (b) O, cA1, cA2, cA3, cA4.

The most informative frequency range for the first group (Figure 6a) is cD4 (4-8 Hz) the average value of $A_{RKF}$ ($A_{RKF\_mean}$) is equal to 67.1%, while for the rest of the frequency ranges $A_{RKF\_mean} \sim 63\%$. Among the signals of the second group (Figure 6b), the most informative is cA3 (0-8 Hz) with an average value of $A_{RKF\_mean} = 71.4\%$, while signals with the presence of higher-frequency components show a lower value of $A_{RKF\_mean}$: 63.2% for O (0-64 Hz), 62.9% for cA1 (0-32 Hz), 65.8% for cA2 (0-16 Hz). The lower accuracy of $A_{RKF\_mean} = 68.2\%$ for cA4 (0-4 Hz) may indicate that the 4-8 Hz range is needed to improve signal classification accuracy. The highest classification accuracy by one feature was obtained for the T8 channel and the cA3 signal: $A_{RKF} = 79.5\%$.

To determine the most informative combinations of channels and frequency ranges, Table 2 was compiled, which contains 15 combinations of channel and signal type with the highest $A_{RKF}$ value from those presented in Figures 6a and 6b. It can be noted that for most of the channels presented in the table (T8, O2, FC6, F3, AF4), only the low-frequency components of the original signal are the most informative: cA3 (0-8 Hz), cA4 (0-4 Hz), cD4 (4- 8 Hz), while for channels F8 and O1, signals with high-frequency components are also informative: O (0-64 Hz) and cD1 (32-64 Hz). It is also worth noting that most of the channels that give the best results were located in the right hemisphere of the head.

Table 2. Combinations of channels and signal type that give the highest $A_{RKF}$ value.

| Channel | Signal type | $A_{RKF}$, % |
|---|---|---|
| T8 | cA3 | 79.5 |
| O1 | cA4 | 77.1 |
| FC6 | cA4 | 76.9 |
| O2 | cA3 | 76.5 |
| FC6 | cA3 | 76.2 |
| F8 | cA2 | 74.9 |
| T8 | cA4 | 74.2 |
| F3 | cA3 | 74.2 |
| F8 | O | 73.4 |
| F8 | cD1 | 73.4 |
| O1 | cA3 | 72.4 |
| F8 | cA3 | 72.3 |
| AF4 | cA3 | 72.1 |
| O1 | O | 71.9 |
| AF4 | cD4 | 71.6 |

**4. Smart IoT environment concept for patient health monitoring**

Based on the results presented in section 3, it is clear that entropy features can be used to analyze EEG signals in order to diagnose Parkinson's disease effectively. Here, we present the idea of a smart IoT environment that can continuously monitor a patient's condition at home (Figure 1). In the presented concept, there are three types of smart ML sensors: one that measures and analyzes EEG signals, another that serves as a video camera sensor, and a third that acts as a cloud IoT gateway. These sensors are able to solve several problems:

1. Determine the effectiveness of the selected treatment methods and monitor the evolution of the disease continuously.
2. Provide the results of the analysis directly to the medical institution without requiring daily face-to-face meetings.
3. Advise medical institutions and/or emergency services that something dangerous/abnormal has occurred.

Continuous monitoring of the patient's condition includes regular (e.g. weekly) EEG measurements at rest and continuous monitoring of motor activity using smart cameras. By analyzing the video image, it is possible to identify specific motor activity disorders characteristic of Parkinson's disease. Both the patient and his/her attending physician can monitor the patient's condition objectively based on the analysis results. Interaction between a smart IoT environment and a medical information system can be achieved through network interaction. Especially relevant are remote northern regions with low population density and long distances to medical institutions with the necessary infrastructure. Additionally, it reduces the burden on medical facilities and reduces the cost and time of transporting patients.

According to the concept of personal medicine [7], constant monitoring of the disease and identifying the best treatment method for everyone are important elements of care. The previous sections discussed the classification of EEG signals used to diagnose Parkinson's disease. FuzzyEn-based features, however, can be used as a tool to assess the current state of a disease. Histograms of entropy values (cA3 for channel T8) for people with Parkinson's disease and healthy controls are shown in Figure 7. Based on the results presented, the presence of disease is associated with more chaotic EEG signals in most patients. Based on the dynamics of the change in the entropy value, it is possible to track the improvement or deterioration of the clinical picture for each individual patient using several combinations of signal type and channel as indicators. As entropy increases, one can speak of deterioration in the patient's condition, and as it decreases, one can speak of improvement. As a result of the variability of values within the dataset under study, the absolute value of entropy cannot serve as an unambiguous indicator of disease severity. The effectiveness of an individual treatment method can also be assessed based on how much entropy has decreased over time compared with control indicators.

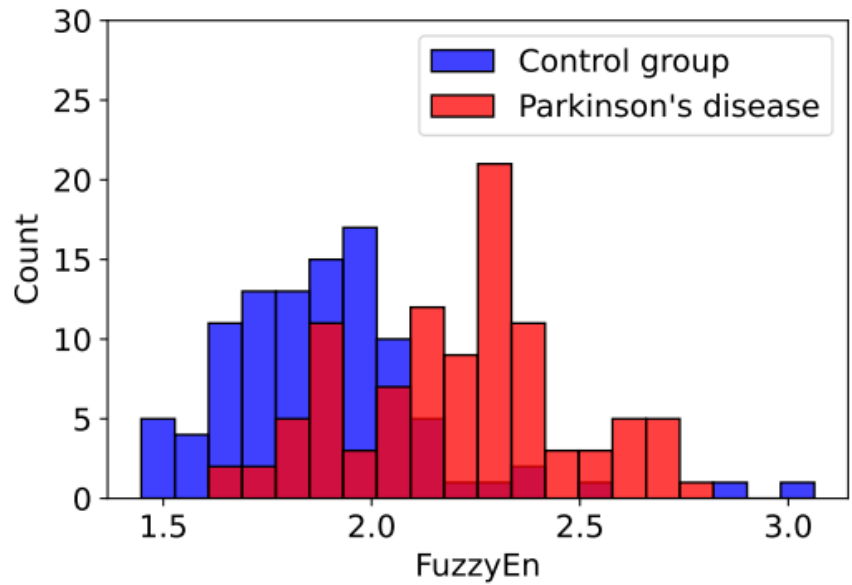

**Figure 7.** Histogram of distribution of FuzzyEn values for signal cA3 of channel T8.

A neurodegenerative disease affects a patient's cognitive skills gradually, leading to a gradual loss of independence. Patients who live with relatives can control their behavior and prevent unwanted and dangerous actions. Using the network infrastructure, the smart IoT environment can alert the hospital and emergency services if the patient lives alone and dangerous situations are resulting from the disease. The optimization of information processing processes is an important step in developing an IoT environment. Due to their limited computing capabilities and small amounts of RAM, IoT devices and gateways need to reduce their volume to speed up data processing. We will present options for optimizing EEG signal processing in the following subsections.

*4.1 Reduction in the number of features*

Section 3.4 showed that different types of signals perform best on different channels. A high classification accuracy can be achieved with a minimum number of features, which appears to be an interesting goal. We examined how the accuracy of the $A_{RKF}$ changes with the number of features computed using FuzzyEn (Section 3.4). In order to do this, we used an iterative approach in which only the first feature gave the maximum value of the $A_{RKF}$.

Next, the $A_{RKF}$ value is calculated for the combination of two features. The evaluation procedure is repeated with one more of the remaining features added. Figure 8 illustrates the dependence of $A_{RKF}$ on feature numbers.

With 11 features, the classification accuracy of the $A_{RKF}$ is 99.9%, which is the same as using all 126 features. It is possible to reduce the number of features to reduce the computational costs of classification and to use lower-performance devices for analysis, such as peripheral IoT devices or embedded analytical modules, by minimizing the number of features, such as peripheral IoT devices or embed analytical modules in EEG signal measurement devices.

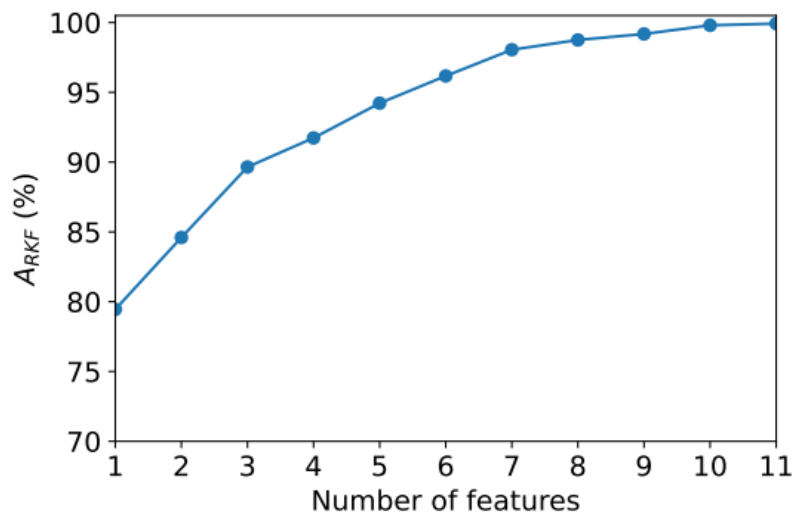

**Figure 8.** Dependence of $A_{RKF}$ on the number of features.

*4.2 Reduction in the segment's length*

The length of the EEG change segment ($L_{EEG}$) can also be reduced to reduce the amount of data to be processed. In section 3, we used segments with 1000 counts (~7.8 seconds). However, it is possible to shorten this length in order to speed up calculations. We will achieve this by reducing the most resource-intensive part of the analysis - the calculation of FuzzyEn entropy. Another part of the time is spent filtering the signal using wavelet methods. According to Figure 9, the $A_{RKF}$ accuracy depends on the number of $L_{EEG}$ readings when using all 126 features (see section 3.1) or the 11 most informative ones (see section 4.1).

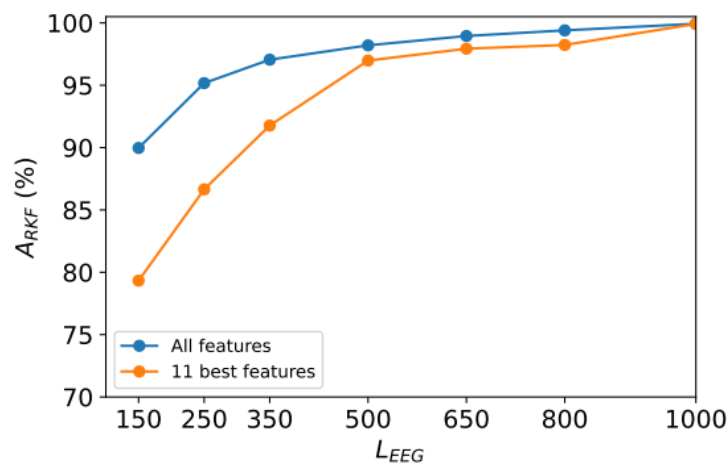

**Figure 9.** Dependence of $A_{RKF}$ on $L_{EEG}$.

As the length of the $L_{EEG}$ segment decreases, the accuracy of the $A_{RKF}$ classification also decreases, but less intensely for 126 features than for 11. In the beginning, the segment length of 1000 samples were chosen as it allowed a high classification accuracy of 99.9%, and at the same time, a decrease in length even by 20% (up to $L_{EEG}$ = 800) resulted in an accuracy decrease of 99.4%, or five times greater $E_{RKF}$ errors. Since the main idea of reducing computational costs is to reduce processing time, we compared the processing time of one segment (calculation of entropy features and classification by the trained model) for different lengths of $L_{EEG}$ segments and different number of features. The calculations were performed on a desktop computer with an Intel i5-7200U (2.5 GHz) processor and 8 GB of RAM.

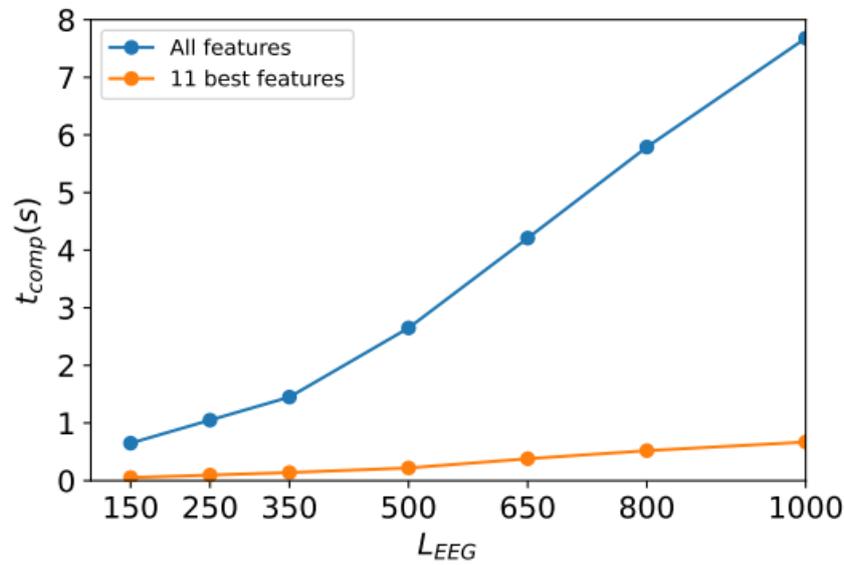

**Figure 10.** Dependence of $t_{comp}$ on $L_{EEG}$.

With more than 350 samples, the processing time $t_{comp}$ depends linearly on the length of $L_{EEG}$ segments, since most of the time is spent calculating entropy features. It took approximately 0.06 seconds to calculate 1 feature with a length of $L_{EEG}$ = 1000. In Figure 11, it can be observed that by reducing the number of features, the calculation time can be significantly reduced (for example, with $L_{EEG}$ = 1000, the calculation time varies by 11 times) while maintaining a low classification error. The reduction of segment length does not significantly improve the calculation speed (for example, the speed difference between $L_{EEG}$ = 1000 and $L_{EEG}$ = 800 is only 25%), but significantly increases the classification error $E_{RKF}$.

## 5. Conclusions

This study proposes a machine learning model based on EEG-entropy features for diagnosing and monitoring Parkinson's disease (PD) in smart IoT environments. We investigated the most effective entropy method to calculate EEG entropy features. We found that fuzzy entropy performed well in detecting and monitoring Parkinson's disease. EEG signals with low frequencies (0-4 Hz) contribute the most to high classification accuracy, and we identified the most prominent EEG signal frequency range. Additionally, the most informative signals were received primarily from the right hemisphere of the head (F8, P8, T8, FC6). A combination of signal frequency range and channels was selected to accurately diagnose the PD with only 11 features achieving the classification accuracy $A_{RKF}$ ~99.9%. The length of the EEG segments ($L_{EEG}$) can also be reduced to reduce the amount of data to be processed. At first, the segment length of 1000 samples is quite optimal since it allows high classification accuracy $A_{RKF}$ = 99.9% for 11 features. A decrease in the length up to $L_{EEG}$ = 800 (~6.2 seconds) leads to a decrease in accuracy to $A_{RKF}$ = 98.3%, and $L_{EEG}$ = 150 (~1.2 seconds) leads to a decrease in accuracy to $A_{RKF}$ = 79.3%. By reducing the segment length and the number of features, computational costs can be reduced, and lower-performance devices can be used in smart IoT environments for ML sensors.

**Author Contributions:** Conceptualization, M.A., A.V. and D.K.; methodology, M.A and A.V.; software, M.A.; validation, M.A., M.M. and A.V.; formal analysis, M.A. and M.M.; investigation, M.A. and M.M.; resources, M.M.; data curation, M.M.; writing—original draft preparation, M.A, M.M., A.V. and D.K.; writing—review and editing, M.A, M.M., A.V. and D.K.; visualization, M.A. and V.A.; supervision, M.M. and D.K.; project administration, D.K.; funding acquisition, D.K. All authors have read and agreed to the published version of the manuscript.

**Funding:** The research is implemented with financial support by Russian Science Foundation, project no. 22-11-20040 (https://rscf.ru/en/project/22-11-20040/) jointly with Republic of Karelia and funding from Venture Investment Fund of Republic of Karelia (VIF RK).

**Institutional Review Board Statement:** The EEG data used in the present study was collected at Hospital UKM medical center in Kuala Lumpur, Malaysia. The ethics statement was obtained from the University Kebangsaan Malaysia (UKM) medical center, Malaysia ethics committee for human research (Ref. number: UKM1.5.3.5/244/FF-354-2012).

**Informed Consent Statement:** Informed consent was obtained from all subjects involved in the study

**Data Availability Statement:** The data are not publicly available due to their containing information that could compromise the privacy of research participants. Data requests can be sent to Murugappan Murugappan through his email m.murugappan@kcst.edu.kw.

**Acknowledgments:** Special thanks to the editors of the journal and to the anonymous reviewers for their constructive criticism and improvement suggestions.

**Conflicts of Interest:** The authors declare no conflict of interest.

FOR PEER REVIEW    19 of 19